\documentclass[10pt,conference]{IEEEtran}\IEEEoverridecommandlockouts
\usepackage{epsfig,graphicx,subfigure,psfrag,amsmath,cases}
\usepackage{latexsym,amssymb,amsmath,epsfig,subfigure,algorithm,mathtools}
\usepackage{algorithmic}
\usepackage{color}
\usepackage{url}
\usepackage{scrtime}

\usepackage{hyperref}
\usepackage{subfigure}
\usepackage{bbding}
\usepackage{multicol}

\author{Zhiqiang Wei,  Linglong Dai, Derrick Wing Kwan Ng, and Jinhong Yuan
\thanks{Zhiqiang Wei, Derrick Wing Kwan Ng, and Jinhong Yuan are with the School of Electrical
Engineering and Telecommunications, the University of New South Wales, Australia (email: zhiqiang.wei@student.unsw.edu.au; w.k.ng@unsw.edu.au; j.yuan@unsw.edu.au).
Linglong Dai is with the Department of
Electronic Engineering, Tsinghua University, China (email: daill@tsinghua.edu.cn). Derrick is supported under Australian Research Council's Scheme Discovery Early Career Researcher Award funding scheme (project
number DE170100137).}
}
\title{Performance Analysis of a Hybrid Downlink-Uplink Cooperative NOMA Scheme}

\newtheorem{Lem}{Lemma}

\newtheorem{T-Prob}{Transformed Problem}

\newcounter{mytempeqncnt}

\newtheorem{Remark}{Remark}

\newcommand{\abs}[1]{\lvert#1\rvert}

\begin{document}
\IEEEspecialpapernotice{(Invited Paper)}
\maketitle

\begin{abstract}
This paper proposes a novel hybrid downlink-uplink cooperative NOMA (HDU-CNOMA) scheme to achieve a better tradeoff between spectral efficiency and signal reception reliability than the conventional cooperative NOMA schemes.  In particular, the proposed scheme enables the strong user to perform a cooperative transmission and an interference-free uplink transmission simultaneously  during the cooperative phase, at the expense of a slightly decrease in signal reception reliability at the weak user. We analyze the outage probability, diversity order, and outage throughput of the proposed scheme. Simulation results not only confirm the accuracy of the developed analytical results, but also unveil  the spectral efficiency gains achieved by the proposed scheme over a baseline cooperative NOMA scheme and a non-cooperative NOMA scheme.
\end{abstract}
\renewcommand{\baselinestretch}{0.97}
\large\normalsize

\section{Introduction}

Recently, non-orthogonal multiple access (NOMA) has drawn a lot of attentions as an important enabling technique to fulfill the challenging requirements of the fifth-generation (5G) communication systems, such as massive connectivity, high spectral efficiency, and ultra-low latency \cite{Kwan:book_2017,Dai2015,WeiSurvey2016}. In the
literature, different schemes, such as power domain NOMA
and code domain NOMA, have been proposed to facilitate
multiuser multiplexing \cite{WeiSurvey2016}.  Power domain NOMA is particularly appealing as it can be integrated with the existing fourth-generation communication systems.
The fundamental idea of power domain NOMA is to exploit the power domain for   multiuser multiplexing via using superposition coding at transmitters and successive interference cancellation (SIC) at receivers \cite{Song2016resource}.
In particular, NOMA allows a strong user (with better channel condition) concurrently accessing the spectrum resources  assigned for a weak user (with worse channel condition) to increase the system spectral efficiency.
To alleviate the inter-user interference (IUI) at the weak user, a larger amount of power is allocated to the weak user while a smaller fraction of power is provided for the strong user. Meanwhile, SIC technique is adopted at the receiver of the strong user to remove the IUI. It has been shown that NOMA provides substantial performance gains over conventional orthogonal multiple access (OMA) in terms of spectral efficiency \cite{Ding2014,Yang2016,Sun2016Fullduplex} and fairness \cite{Timotheou2015,LiuFairnessNOMA}.

In wireless communications, the system performance is significantly limited by channel fading raised from multipath propagations.
This issue is more prominent in NOMA scenarios. Specifically, weak users become more vulnerable to channel fading due to not only the severe path loss, but also the IUI caused by the simultaneous communication to strong users.
Traditionally, cooperative diversity is an effective technique to combat channel fadings in wireless networks \cite{LanemanCoopDiveristy}. Among different cooperative strategies proposed in the literature \cite{JR:cooperative_xiaoming, Zhang:15,JR:Cooperative}, cooperative relaying is an attractive technique to increase the range of communication systems and to enhance
the link reliability without incurring the high cost of additional
base station deployment. Therefore, a cooperative NOMA (CNOMA) scheme was proposed in \cite{Ding2015} to improve the signal reception reliability for the weak user by exploiting the prior information obtained at the strong user during SIC process.
Particularly, in addition to the downlink NOMA transmission phase, the strong user acts as a decode-and-forward (DF) relay to deliver messages to the weak user in the cooperative phase.
The extensions of this scheme to multiple-antenna relaying networks and full-duplex relaying networks were investigated in \cite{Men2015NOMA} and \cite{ZhangD2DNOMA}, respectively.
Note that the aforementioned CNOMA schemes enhance the signal reception reliability at the price of reduced spectral efficiency due to the duplicate transmission during the cooperative phase.
%Besides, the idea of simultaneous wireless information and power transfer (SWIPT) was introduced to cooperative NOMA systems, in which the strong user serves as an energy harvesting relay to assist the weak user utilizing the harvested energy from radio frequency.
More recently, a non-orthogonal relaying strategy is applied in CNOMA systems to improve the spectral efficiency, where a base station (BS) and a relay transmit their messages at the same time in the same frequency.
Nevertheless, a dedicated relay is required in most of existing schemes \cite{KimCDRT2015,ZhongFullDuplex}.
Also, these schemes do not fully exploit the BS in the cooperative phase which lead to potential loss in spectral efficiency.
%In addition, the side information at the BS is very promising to achieve a higher spectral efficiency if we combine the uplink and downlink transmission in an effective way, which has never been exploited in existing works about cooperative NOMA.
%The further extension to full-duplex relaying networks was presented in \cite{ZhongFullDuplex}
%and the optimal relay selection strategy was proposed in \cite{Ding2016a}.

In this paper, we propose a new hybrid downlink-uplink CNOMA (HDU-CNOMA) scheme to improve the spectral efficiency.
Different from the conventional CNOMA scheme\cite{Ding2015}, our proposed scheme enables the uplink transmission from the strong user to the BS during the cooperative phase.
%Since the cooperative phase is utilized for downlink transmission of the weak user and uplink transmission of the strong user simultaneously, our proposed scheme is named as hybrid downlink-uplink NOMA.
Hence, it is expected that our proposed scheme is able to improve the achievable system sum rate at a price of a slightly decrease in the signal reception reliability at the  weak user.
%Note that the BS exactly knows what it has sent to the weak user in advance, the superposition transmission in the cooperative phase will not affect the uplink transmission of the strong user except for some power loss.
%Therefore,
Besides, we derive the closed-form expressions of the system outage probability and the diversity orders to characterize the performance of the proposed scheme.
Numerical results are shown to verify our analytical results and to demonstrate the effectiveness of our proposed scheme.
%Compared to non-cooperative NOMA scheme and conventional CNOMA scheme, our proposed scheme can achieve a significant improvement in spectral efficiency.

%Notations used in this paper are as follows. Boldface capital and lower case letters are reserved for matrices and vectors, respectively, ${\left( \cdot \right)^T}$ denotes the transpose of a vector or matrix.
%$\mathbb{C}^{M\times N}$ denotes the set of all $M\times N$ matrices with complex entries; $\mathbb{R}^{M\times N}$ denotes the set of all $M\times N$ matrices with real entries;
%$\abs{\cdot}$ denotes the absolute value of a complex scalar; $\left\| \cdot \right\|_2$ denotes the $l_2$-norm of a vector; and
%$\Pr \left\{  \cdot  \right\}$ denotes the probability of a random event; ${E_x}\left\{  \cdot  \right\}$ denotes the expectation with respective to (w.r.t.) a random variable $x$.
%The circularly symmetric complex Gaussian distribution with mean $\mu$ and variance $\sigma^2$ is denoted by ${\cal CN}(\mu,\sigma^2)$;
%$\sim$ stands for ``distributed as"; $U[a,b]$ denotes the uniform distribution in the interval $[a, b]$; and
%${\nabla _{\mathbf{x}}}f$ denotes the gradient of a function $f$ with respective to vector ${\mathbf{x}}$.

%\begin{figure}[t]
%\centering
%\includegraphics[width=4in]{ConventionalCooperativeNOMAModelSingleFrame.eps}
%\caption{Conventional CNOMA scheme with one base station and two users\cite{Ding2015}.}
%\label{CooperativeNOMAModel}
%\end{figure}

\begin{figure}[t]
\centering
\includegraphics[width=3.5in]{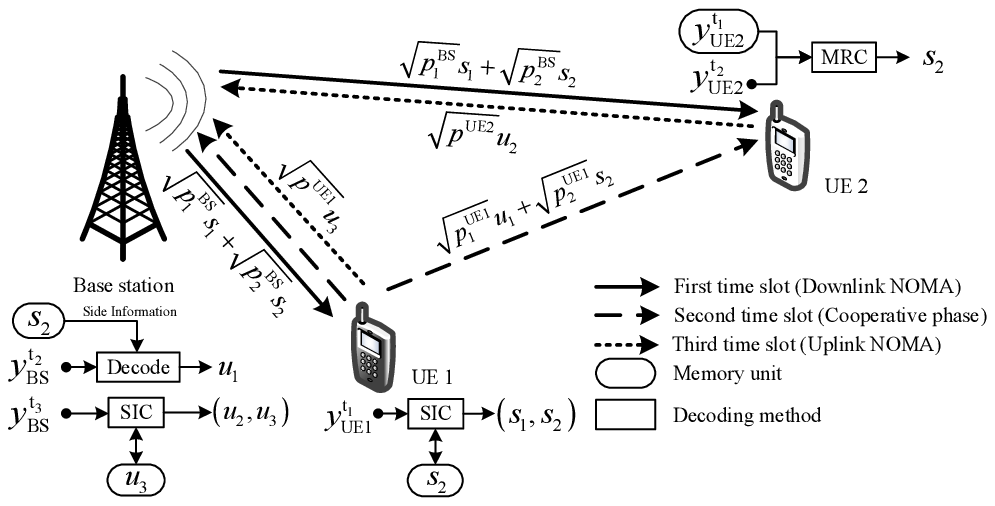}
\caption{The proposed HDU-CNOMA scheme with one BS and two users.}
\label{HybridCooperativeNOMAModel}
\end{figure}

%\begin{figure}[t]
%\centering\vspace*{-7mm}
%\subfigure[Outage probability for all the users with $\kappa^2_{i,m} = 0.1$.]
%{\label{CaseV_OutageProbability:a} %% label for first subfigure
%\includegraphics[width=0.46\textwidth]{CaseV_OutageProbability_1.eps}}\vspace*{-1mm}
%\subfigure[Outage probability versus $\kappa^2_{i,m}$ for user 9.]
%{\label{CaseV_OutageProbability:b} %% label for second subfigure
%\includegraphics[width=0.465\textwidth]{CaseV_OutageProbability_2.eps}}\vspace*{-1mm}
%\caption{Outage probability of our proposed scheme and a naive scheme with $N_\mathrm{F} = 8$ and $M = 12$.}\vspace*{-9mm}
%\label{CaseV_OutageProbability}%
%\end{figure}
\section{System Model}
%\textcolor[rgb]{0.00,0.07,1.00}{\subsection{Main Assumption}
%(List all the main assumptions as follows.)
%\begin{itemize}
%  \item All transceivers are equipped with a single antenna and work in half-duplex mode.
%  \item Time division duplex (TDD) for uplink and downlink transmissions.
%  \item Ability of storage for base station (BS), user 1 (UE 1), and user 2 (UE 2).
%  \item Ability of superposition coding for UE 1.
%  \item Perfect channel state information (CSI) is available for detection at the receivers, while only statistical CSI is available for transmission design at the transmitters.
%  \item Quasi-static Rayleigh fading, the CSI will not change during one frame.
%\end{itemize}}
Consider a communication scenario including downlink and uplink transmission with one BS and two users\footnote{The extension to the case with more than two users is straightforward by following a similar approach as \cite{Ding2015}.}, as shown in Figure \ref{HybridCooperativeNOMAModel}.
All the transceivers are equipped with a single antenna and operate in half-duplex mode, i.e., they cannot transmit and receive a signal at the same time in the same frequency.
Furthermore, we assume a time division duplex (TDD) protocol for facilitating downlink and uplink transmission.
We denote $h_{\mathrm{BS,UE 1}}$ as the channel coefficient between the BS and user 1 (UE 1), $h_{\mathrm{BS,UE 2}}$ as the channel coefficient between the BS and user 2 (UE 2), and $h_{\mathrm{UE 1,UE 2}}$ as the channel coefficient between UE 1 and UE 2. We assume that perfect channel state information (CSI) is available at receivers for signal detection, while only statistical CSI is available at transmitters. All the links considered here are assumed to experience independent quasi-static fading, where the channel coefficients are constant for each time slot but vary independently between different time slots for different links.
 Besides, we assume that the channel coefficients are Rayleigh  distributed: $h_{\delta} \sim {\cal CN}(0,\beta_{\delta})$, $\delta \in \left\{ \mathrm{(BS,UE 1), (BS,UE 2),(UE 1,UE 2)} \right\}$, where ${\cal CN}(0,\beta_{\delta})$ denotes the circularly symmetric complex Gaussian distribution with zero-mean and variance $\beta_{\delta}$, and the variance $\beta_{\delta}$ captures the effect of large scale fading for the link $\delta$.  Then, the cumulative distribution function (CDF) and probability density function (PDF) for the channel gain of link $\delta$, i.e., ${{\left| {{{{h}}_{{\delta}}}} \right|}^2}$, are given by
%\vspace*{-1.5mm}
\begin{align}
\hspace*{-1mm}{F_{{{\left| {{{{h}}_{{\delta}}}} \right|}^2}}}\left(x\right) &= 1 - \exp(-\frac{x}{{\beta _{{\delta}}}}),\; x\ge 0,\;\text{and}\\
\hspace*{-1mm}{f_{{{\left| {{{\mathrm{h}}_{{\delta}}}} \right|}^2}}}\left(x\right) &= \frac{1}{{\beta _{{\delta}}}}\exp(-\frac{x}{{\beta _{{\delta}}}}),\; x\ge 0,
\end{align}
%\par
%\vspace*{-1.5mm}
%\noindent
respectively, where $\abs{\cdot}$ denotes the absolute value of a complex scalar.
Meanwhile, we consider the user with the larger $\beta_{\delta}$ as the strong user and without loss of generality, we assume $\beta_{\mathrm{BS,UE 1}} > \beta_{\mathrm{BS,UE 2}}$.
In other words, UE 1 is selected to perform SIC and to assist UE 2 in our proposed scheme\cite{CuiPowerEfficientNOMA,KimCooperativeNOMA2015}.
Note that this may not be the optimal SIC decoding order to minimize the system outage probability under statistical CSI assumption\cite{CuiPowerEfficientNOMA,Wei2016NOMA}, because $\beta_{\mathrm{BS,UE 1}} > \beta_{\mathrm{BS,UE 2}}$ does not guarantee ${\left| {h_{\mathrm{BS,UE 1}}} \right|^2} > {\left| {h_{\mathrm{BS,UE 2}}} \right|^2}$.
However, it is a simple but effective strategy under statistical CSI\cite{CuiPowerEfficientNOMA}.
To facilitate our performance analysis, we focus on this specific scheme with UE 1 as the strong user and serving as a relay to assist UE 2.

\begin{figure}[t]
\centering
\subfigure[Proposed HDU-CNOMA scheme.]
{\label{TransmissionIllustration:a} %% label for second subfigure
\includegraphics[width=0.35\textwidth]{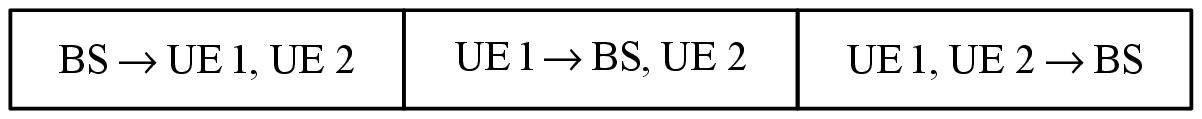}}\vspace*{-2mm}
\subfigure[Conventional CNOMA scheme\cite{Ding2015}.]
{\label{TransmissionIllustration:b} %% label for first subfigure
\includegraphics[width=0.35\textwidth]{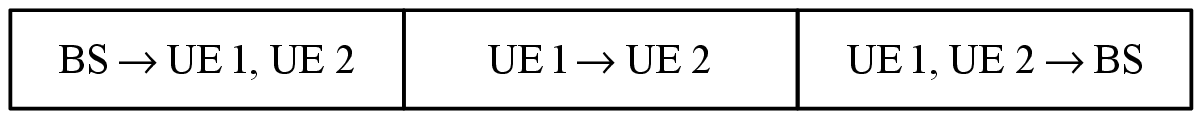}}\vspace*{-2mm}
\subfigure[Non-cooperative NOMA scheme.]
{\label{TransmissionIllustration:c} %% label for first subfigure
\includegraphics[width=0.35\textwidth]{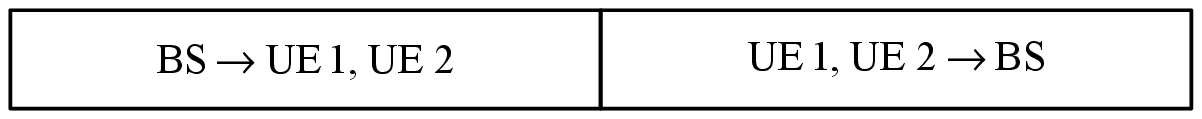}}\vspace*{-2mm}
%\subfigure[Non-cooperative OMA scheme.]
%{\label{TransmissionIllustration:d} %% label for first subfigure
%\includegraphics[width=0.35\textwidth]{TimeSlots_OMA.eps}}%\vspace*{-2mm}
\caption{Illustrations for: a) proposed HDU-CNOMA scheme; b) conventional CNOMA scheme\cite{Ding2015}; c) non-cooperative NOMA scheme.}
\label{TransmissionIllustration}%
\end{figure}

As shown in Figure \ref{HybridCooperativeNOMAModel} and Figure \ref{TransmissionIllustration:a}, in our proposed HDU-CNOMA scheme, one time frame is partitioned into three time slots with equal duration  for downlink NOMA phase, cooperative phase, and uplink NOMA phase.
Note that fixed power allocation is adopted for in this paper. Although optimizing the power allocation during different phases can further improve the performance of our proposed scheme, it is beyond the scope of this paper and will be considered in our future work. In the following, we present our proposed scheme.

\subsection{Proposed HDU-CNOMA Scheme}
In the first time slot, i.e.,  the downlink NOMA phase, the transmitted signal from the BS is given by
%\vspace*{-1mm}
\begin{equation}
x_{{\mathrm{BS}}}^{{\mathrm{t_1}}} = \sqrt {\alpha_{{\mathrm{UE1}}}^{{\mathrm{t_1}}} P_0} {{s_1}} + \sqrt {\alpha_{{\mathrm{UE2}}}^{{\mathrm{t_1}}} P_0} {{s_2}},
%\vspace*{-1mm}
\end{equation}
where superscript $\mathrm{t_1}$ denotes the 1-st time slot, $P_0$ denotes the maximum transmit power for the BS, ${{s_1}}$ and ${{s_2}}$ denote the modulated downlink symbols for UE 1 and UE 2, respectively, and $\alpha_{{\mathrm{UE1}}}^{{\mathrm{t_1}}}$ and $\alpha_{{\mathrm{UE2}}}^{{\mathrm{t_1}}}$ denote the power allocation factors for UE 1 and UE 2 in $\mathrm{t_1}$, respectively.
According to the NOMA protocol\cite{Ding2014}, we allocate more power to the weak user, thus we have $\alpha_{{\mathrm{UE1}}}^{{\mathrm{t_1}}} \le \alpha_{{\mathrm{UE2}}}^{{\mathrm{t_1}}}$ and $\alpha_{{\mathrm{UE1}}}^{{\mathrm{t_1}}} + \alpha_{{\mathrm{UE2}}}^{{\mathrm{t_1}}} = 1$.
For notational simplicity, we assume the same maximum transmit power for the BS, UE 1, and UE 2 in our model\footnote{{Note that it is straightforward to extend the results of this paper to the case with different transmit powers.}}.
Subsequently, the received signals at UE 1 and UE 2 in $\mathrm{t_1}$ are given by
%\vspace*{-2mm}
\begin{align}
\hspace*{-2mm}y_{{\mathrm{UE 1}}}^{{\mathrm{t_1}}} &= {h_{{\mathrm{BS}},{\mathrm{UE 1}}}}\left( \hspace*{-1mm}\sqrt {\alpha_{{\mathrm{UE1}}}^{{\mathrm{t_1}}} P_0} {{s_1}} \hspace*{-1mm}+\hspace*{-1mm} \sqrt {\alpha_{{\mathrm{UE2}}}^{{\mathrm{t_1}}} P_0} {{s_2}} \hspace*{-1mm}\right) \hspace*{-1mm}+\hspace*{-1mm} {z_{{\mathrm{UE 1}}}}\; \text{and} \\[-1mm]
\hspace*{-2mm}y_{{\mathrm{UE 2}}}^{{\mathrm{t_1}}} &= {h_{{\mathrm{BS}},{\mathrm{UE 2}}}}\left(\hspace*{-1mm} \sqrt {\alpha_{{\mathrm{UE1}}}^{{\mathrm{t_1}}} P_0} {{s_1}} \hspace*{-1mm}+\hspace*{-1mm} \sqrt {\alpha_{{\mathrm{UE2}}}^{{\mathrm{t_1}}} P_0} {{s_2}} \hspace*{-1mm}\right) \hspace*{-1mm}+\hspace*{-1mm} {z_{{\mathrm{UE 2}}}},
\end{align}
%\par
%\vspace*{-2mm}
%\noindent
respectively, where ${z_{{\mathrm{UE 1}}}}\sim{\cal CN}(0,\sigma^2)$ and ${z_{{\mathrm{UE 2}}}}\sim{\cal CN}(0,\sigma^2)$ denote the additive white Gaussian noise (AWGN) at UE 1 and UE 2, respectively, with the same noise power $\sigma^2$.

Then, UE 1 will first decode message of UE 2 $s_2$, subtract it from its observation $y_{{\mathrm{UE 1}}}^{{\mathrm{t_1}}}$, and then decode its own message $s_1$.
The signal-to-interference-plus-noise ratio (SINR) for UE 1 to decode the message of UE 2 is given by
%\vspace*{-2.5mm}
\begin{equation}
{\mathrm{SINR}}_{{\mathrm{UE 1,UE 2}}}^{{\mathrm{t_1}}} = \frac{{{{\left| {{h_{{\mathrm{BS}},{\mathrm{UE 1}}}}} \right|}^2}{\alpha_{{\mathrm{UE2}}}^{{\mathrm{t_1}}}}}}{{{{\left| {{h_{{\mathrm{BS}},{\mathrm{UE 1}}}}} \right|}^2}{\alpha_{{\mathrm{UE1}}}^{{\mathrm{t_1}}}} + 1/\rho}}.
%\vspace*{-2mm}
\end{equation}
where $\rho = \frac{P_0}{\sigma^2}$ denotes the transmit signal-to-noise ratio (SNR).
For a given target data rate of  downlink transmission of UE 2, ${{R}_{{\mathrm{UE 2}}}^{\mathrm{DL}}}$, if $\frac{1}{3}{\log _2}\left( {1 + {\mathrm{SINR}}_{{\mathrm{UE 1,UE 2}}}^{{\mathrm{t_1}}}} \right) \ge {{R}_{{\mathrm{UE 2}}}^{\mathrm{DL}}}$, the message $s_2$ is decodable and can be cancelled at UE 1, otherwise the SIC process is failed.
Note that a pre-log factor of $\frac{1}{3}$ is introduced which takes into account the loss of spectral efficiency as one time frame is partitioned into three time slots.
Meanwhile, we assume that UE 1 will not decode its own message $s_1$ if the SIC process is failed.
Therefore, with a successful SIC, the SINR for UE 1 to decode its own messages is given by
%\vspace*{-1mm}
\begin{equation}
{\mathrm{SINR}}_{{\mathrm{UE 1}}}^{{\mathrm{t_1}}} = {{{\left| {{h_{{\mathrm{BS}},{\mathrm{UE 1}}}}} \right|}^2}{\alpha_{{\mathrm{UE1}}}^{{\mathrm{t_1}}}}}\rho.
%\vspace*{-1mm}
\end{equation}

On the other hand, UE 2 will directly decode its own message $s_2$ by treating the signal of UE 1 as noise. Thereby, the SINR for UE 2 to decode its own message is given by
\begin{equation}
{\mathrm{SINR}}_{{\mathrm{UE 2}}}^{{\mathrm{t_1}}} = \frac{{{{\left| {{h_{{\mathrm{BS}},{\mathrm{UE 2}}}}} \right|}^2}{\alpha_{{\mathrm{UE2}}}^{{\mathrm{t_1}}}}}}{{{{\left| {{h_{{\mathrm{BS}},{\mathrm{UE 2}}}}} \right|}^2}{\alpha_{{\mathrm{UE1}}}^{{\mathrm{t_1}}}} + 1/\rho}}.
%\vspace*{-2mm}
\end{equation}
%Note that ${\mathrm{SINR}}_{{\mathrm{UE 2,UE 2}}}^{{\mathrm{t_1}}} \le {\mathrm{SINR}}_{{\mathrm{UE 1,UE 2}}}^{{\mathrm{t_1}}}$ if $\left|h_{\mathrm{BS,UE 1}}\right| \ge \left|h_{\mathrm{BS,UE 2}}\right|$, which means that UE 1 can decode the message of UE 2 as long as UE 2 can decode its own message in the first time slot.

In the second time slot $\mathrm{t_2}$, i.e., the cooperative phase,  UE 1 will broadcast the superimposed signal of ${s}_2$ and $u_1$, where ${s}_2$ is the message for UE 2 obtained during SIC process in the first time slot and $u_1$ is its own message for uplink transmission.
The transmitted signal from UE 1 in the second time slot is given by
%\vspace*{-1mm}
\begin{equation}
x_{{\mathrm{UE 1}}}^{{\mathrm{t_2}}} = \sqrt {\alpha_{{\mathrm{BS}}}^{{\mathrm{t_2}}} P_0} {u_1} + \sqrt {\alpha_{{\mathrm{UE2}}}^{{\mathrm{t_2}}} P_0} {{s}_2},
%\vspace*{-1mm}
\end{equation}
where $\alpha_{{\mathrm{BS}}}^{{\mathrm{t_2}}}$ and $\alpha_{{\mathrm{UE2}}}^{{\mathrm{t_2}}}$ denote the power allocation factors for the messages for the BS and UE 2 in $\mathrm{t_2}$, respectively, with $\alpha_{{\mathrm{BS}}}^{{\mathrm{t_2}}} + \alpha_{{\mathrm{UE2}}}^{{\mathrm{t_2}}} = 1$.
As a result, the received signal at the BS and UE 2 in ${\mathrm{t_2}}$ are given by
%\vspace*{-3mm}
\begin{align}
y_{{\mathrm{BS}}}^{{\mathrm{t_2}}} &= {h_{{\mathrm{BS}},{\mathrm{UE 1}}}}\left(\hspace*{-1mm} \sqrt {\alpha_{{\mathrm{BS}}}^{{\mathrm{t_2}}} P_0} {u_1} \hspace*{-1mm}+\hspace*{-1mm} \sqrt {\alpha_{{\mathrm{UE2}}}^{{\mathrm{t_2}}} P_0} {{s}_2} \hspace*{-1mm}\right) \hspace*{-1mm}+\hspace*{-1mm} {z_{{\mathrm{BS}}}}\; \text{and} \\%[-1mm]
y_{{\mathrm{UE 2}}}^{{\mathrm{t_2}}} &= {h_{{\mathrm{UE 1}},{\mathrm{UE 2}}}}\left( \hspace*{-1mm}\sqrt {\alpha_{{\mathrm{BS}}}^{{\mathrm{t_2}}} P_0} {u_1} \hspace*{-1mm}+\hspace*{-1mm} \sqrt {\alpha_{{\mathrm{UE2}}}^{{\mathrm{t_2}}} P_0} {{s}_2} \hspace*{-1mm}\right)\hspace*{-1mm} +\hspace*{-1mm} {z_{{\mathrm{UE 2}}}},
\end{align}
%\par
%\vspace*{-3mm}
%\noindent
respectively, where ${z_{{\mathrm{BS}}}}\sim{\cal CN}(0,\sigma^2)$ denotes the AWGN at the BS.

Since the BS knows exactly the downlink message $s_2$ in advance, it can subtract it directly from its observation $y_{{\mathrm{BS}}}^{{\mathrm{t_2}}}$ and decode the uplink message ${u_1}$.
In other words, the downlink message $s_2$ stored at the BS serves as a piece of side information which benefits the decoding of the uplink message $u_1$.
Therefore, our proposed scheme enables an interference-free uplink transmission and can significantly increase the system spectral efficiency.
On the other hand, compared to the conventional CNOMA scheme, it is expected that there is a slightly decrease in  the signal reception reliability at UE 2 as a portion of transmit power at UE 1,  $\sqrt {\alpha_{{\mathrm{UE2}}}^{{\mathrm{t_2}}} P_0}$, is used for  uplink transmission for UE 1. In fact, allocating a small fraction of power for the uplink transmission of UE 1 can enable a noticeable system performance gain in spectral efficiency owing to its good channel condition and the interference-free transmission. Therefore, in the proposed scheme,  one can use the power allocation factor $\alpha_{{\mathrm{UE2}}}^{{\mathrm{t_2}}}$ to control the tradeoff between system spectral efficiency and signal reception reliability.
Note that  the conventional CNOMA scheme is a subcase of our proposed scheme which can be obtained by setting  $\alpha_{{\mathrm{UE2}}}^{{\mathrm{t_2}}} = 0$.
More importantly, unlike the SIC process in $\mathrm{t_1}$ at UE 1, the downlink message $s_2$ can always be cancelled disregard the target data rate of the downlink transmission of UE 2.

At the BS, after eliminating $s_2$ from $y_{{\mathrm{BS}}}^{{\mathrm{t_2}}}$, the SINR for the BS to decode the uplink message $u_1$ is given by
%\vspace*{-1mm}
\begin{equation}
{\mathrm{SINR}}_{{\mathrm{BS,UE 1}}}^{{\mathrm{t_2}}} = {{{\left| {{h_{{\mathrm{BS}},{\mathrm{UE 1}}}}} \right|}^2}{\alpha_{{\mathrm{BS}}}^{{\mathrm{t_2}}}}}\rho.
%\vspace*{-1mm}
\end{equation}
On the other hand, at UE 2, the maximum ratio combining (MRC) is adopted to decode the message $s_2$ from two independent observations $y_{{\mathrm{UE 2}}}^{{\mathrm{t_1}}}$ and $y_{{\mathrm{UE 2}}}^{{\mathrm{t_2}}}$ with weights $\frac{{h_{{\mathrm{BS}},{\mathrm{UE 2}}}^ * \sqrt {\alpha_{{\mathrm{UE2}}}^{{\mathrm{t_1}}} P_0} }}{{{{\left| {{h_{{\mathrm{BS}},{\mathrm{UE 2}}}}} \right|}^2}{\alpha_{{\mathrm{UE1}}}^{{\mathrm{t_1}}} P_0} + \sigma^2}}$ and $\frac{{h_{{\mathrm{UE 1}},{\mathrm{UE 2}}}^ * \sqrt {\alpha_{{\mathrm{UE2}}}^{{\mathrm{t_2}}} P_0} }}{{{{\left| {{h_{{\mathrm{UE 1}},{\mathrm{UE 2}}}}} \right|}^2}{\alpha_{{\mathrm{BS}}}^{{\mathrm{t_2}}} P_0} + \sigma^2}}$, respectively, where $*$ denotes the conjugate operation. Therefore, the SINR for UE 2 to decode $s_2$ with MRC is given by
%\vspace*{-1mm}
\begin{equation}
{\mathrm{SINR}}_{{\mathrm{UE 2-MRC}}}^{{\mathrm{t_1}},{\mathrm{t_2}}} =
{\mathrm{SINR}}_{{\mathrm{UE 2}}}^{{\mathrm{t_1}}}+ {\mathrm{SINR}}_{{\mathrm{UE 2}}}^{{\mathrm{t_2}}},
%\vspace*{-1mm}
\end{equation}
where ${\mathrm{SINR}}_{{\mathrm{UE 2}}}^{{\mathrm{t_2}}}$ denotes the SINR for UE 2 to decode $s_2$ in $\mathrm{t_2}$, and it is given by
%\vspace*{-1mm}
\begin{equation}
{\mathrm{SINR}}_{{\mathrm{UE 2}}}^{{\mathrm{t_2}}} = \frac{{{{\left| {{h_{{\mathrm{UE 1}},{\mathrm{UE 2}}}}} \right|}^2}{\alpha_{{\mathrm{UE2}}}^{{\mathrm{t_2}}}}}}{{{{\left| {{h_{{\mathrm{UE 1}},{\mathrm{UE 2}}}}} \right|}^2}{\alpha_{{\mathrm{BS}}}^{{\mathrm{t_2}}}} + 1/\rho}}.
%\vspace*{-1mm}
\end{equation}

In the third time slot $\mathrm{t_3}$, i.e., the uplink NOMA phase,  UE 1 and UE 2 transmit their uplink messages $u_3$ and $u_2$ to the BS simultaneously.
Note that the different large scale fading experienced by both users results in different received signal powers from UE 1 and UE 2, which can inherently facilitate the SIC process.
Therefore, we simply assume that both users transmit their messages with their maximum transmit powers for notation simplification.
The received signal at the BS in the third time slot is given by
%\vspace*{-1mm}
\begin{equation}
y_{{\mathrm{BS}}}^{{\mathrm{t_3}}} = {h_{{\mathrm{BS}},{\mathrm{UE 1}}}}\sqrt {P_0} {u_3} + {h_{{\mathrm{BS}},{\mathrm{UE 2}}}}\sqrt {P_0} {u_2} + {z_{{\mathrm{BS}}}}.
%\vspace*{-1mm}
\end{equation}
According to the uplink NOMA principle\cite{Tse2005}, the BS will first decode the user with higher received power. If ${\left| {h_{\mathrm{BS,UE 1}}} \right|^2} \ge {\left| {h_{\mathrm{BS,UE 2}}} \right|^2}$, the SINR for the BS to decode the uplink messages of UE 1 and UE 2 are given by
%\vspace*{-2mm}
\begin{align}
{\mathrm{SINR}}_{{\mathrm{BS,UE1}}}^{{\mathrm{t_3}}} &= \frac{{{{\left| {{h_{{\mathrm{BS}},{\mathrm{UE1}}}}} \right|}^2}}}{{{\left| {{h_{{\mathrm{BS}},{\mathrm{UE2}}}}} \right|}^2} + 1/\rho}\; \;\text{and}  \\%[-1mm]
{\mathrm{SINR}}_{{\mathrm{BS,UE2}}}^{{\mathrm{t_3}}} &= {{{\left| {{h_{{\mathrm{BS}},{\mathrm{UE2}}}}} \right|}^2}}\rho,
\end{align}
%\par
%\vspace*{-2mm}
%\noindent
respectively. On the other hand, if ${\left| {h_{\mathrm{BS,UE 1}}} \right|^2} < {\left| {h_{\mathrm{BS,UE 2}}} \right|^2}$, the SINR for the BS to decode the uplink messages of UE 1 and UE 2 are given by
%\vspace*{-2mm}
\begin{align}
{\mathrm{\overline{SINR}}}_{{\mathrm{BS,UE1}}}^{{\mathrm{t_3}}} &= {{{\left| {{h_{{\mathrm{BS}},{\mathrm{UE1}}}}} \right|}^2}}\rho\;\;\text{and} \\[-1mm] {\mathrm{\overline{SINR}}}_{{\mathrm{BS,UE2}}}^{{\mathrm{t_3}}} &= \frac{{{{\left| {{h_{{\mathrm{BS}},{\mathrm{UE2}}}}} \right|}^2}}}{{{\left| {{h_{{\mathrm{BS}},{\mathrm{UE1}}}}} \right|}^2} + 1/\rho},
\end{align}
%\par
%\vspace*{-2mm}
%\noindent
%\begin{align}
%\hspace*{-3mm}\left( {{\mathrm{SINR}}_{{\mathrm{BS}},{\mathrm{UE1}}}^{{{\mathrm{t}}_{\mathrm{3}}}},{\mathrm{SINR}}_{{\mathrm{BS}},{\mathrm{UE2}}}^{{{\mathrm{t}}_{\mathrm{3}}}}} \right) = \left\{ \begin{array}{l}
%\hspace*{-2mm}\left( {\frac{{{{\left| {{h_{{\mathrm{BS}},{\mathrm{UE1}}}}} \right|}^2}}}{{{{\left| {{h_{{\mathrm{BS}},{\mathrm{UE2}}}}} \right|}^2} + 1/\rho }},{{\left| {{h_{{\mathrm{BS}},{\mathrm{UE2}}}}} \right|}^2}\rho } \right), \;\; \text{if} \;{{{\left| {{h_{{\mathrm{BS}},{\mathrm{UE1}}}}} \right|}^2}} \ge {{{\left| {{h_{{\mathrm{BS}},{\mathrm{UE2}}}}} \right|}^2}},\\
%\hspace*{-2mm}\left( {{{\left| {{h_{{\mathrm{BS}},{\mathrm{UE1}}}}} \right|}^2}\rho ,\frac{{{{\left| {{h_{{\mathrm{BS}},{\mathrm{UE2}}}}} \right|}^2}}}{{{{\left| {{h_{{\mathrm{BS}},{\mathrm{UE1}}}}} \right|}^2} + 1/\rho }}} \right), \;\; \text{otherwise}.
%\end{array} \right.
%\end{align}
respectively. Here, we assume that the BS will not decode the message of the user with lower received power if the SIC process is failed.

%, while non-cooperative OMA scheme, shown in  Figure \ref{TransmissionIllustration:d}, needs four time slots for downlink and uplink OMA transmissions
\begin{Remark}
For comparison, two baseline schemes, the conventional CNOMA scheme and the non-cooperative NOMA scheme, are illustrated in Figure \ref{TransmissionIllustration:b} and Figure \ref{TransmissionIllustration:c}, respectively. For a fair comparison, the time duration of the frame for all the schemes illustrated in Figure \ref{TransmissionIllustration} are identical. Similar to our proposed scheme, the CNOMA scheme also requires three time slots to accomplish the downlink transmission, cooperative transmission, and uplink transmission. Different from the CNOMA scheme, UE 1 in our proposed scheme will broadcast the superposition of downlink symbols for UE 2 and uplink symbols of itself in the cooperative phase. In contrast, the non-cooperative NOMA scheme needs two time slots for downlink NOMA and uplink NOMA transmissions.
\end{Remark}
\section{Performance Analysis}
To characterize the reception reliability and system spectral efficiency of our proposed scheme, three performance metrics are discussed in this section. Firstly, we analyze the outage probability for individual link for a given the target data rate, from which the diversity order achieved by the proposed scheme is obtained. Then, the system outage throughput is derived to demonstrate the improvement of spectral efficiency.

Given the target data rate for downlink and uplink transmissions of UE 1 and UE 2 as ${{R}_{{\mathrm{UE 1}}}^{\mathrm{DL}}}$, ${{R}_{{\mathrm{UE 2}}}^{\mathrm{DL}}}$, ${{R}_{{\mathrm{UE 1}}}^{\mathrm{UL}}}$,
${{R}_{{\mathrm{UE 2}}}^{\mathrm{UL}}}$, respectively, an outage occurs when the achievable rate is less than that of the corresponding target data rate.
Accordingly, the outage probability of downlink and uplink transmissions of UE 1 and UE 2 are given by (\ref{OutageProbability_UE1Down})-(\ref{OutageProbability_UE2UP}) at the top of next page. Note that we assume the same target data rate ${{R}_{{\mathrm{UE 1}}}^{\mathrm{UL}}}$ for the uplink transmissions of UE 1 in $\mathrm{t_2}$ and $\mathrm{t_3}$. Correspondingly, their outage probability are denoted as ${\mathrm{P}}_{{\mathrm{out,\;t_2}}}^{{\mathrm{UE 1,\;UL}}}$ and ${\mathrm{P}}_{{\mathrm{out,\;t_3}}}^{{\mathrm{UE 1,\;UL}}}$, respectively.

\begin{figure*}[!t]
% ensure that we have normalsize text
\normalsize
% Store the current equation number.
\setcounter{mytempeqncnt}{\value{equation}}
% Set the equation number to one less than the one
% desired for the first equation here.
% The value here will have to changed if equations
% are added or removed prior to the place these
% equations are referenced in the main text.
\setcounter{equation}{19}
\begin{align}
\hspace*{-3mm}{\mathrm{P}}_{{\mathrm{out}}}^{{\mathrm{UE 1,\;DL}}} \hspace*{-1mm}=&\hspace*{-0.5mm} \Pr \hspace*{-0.5mm}\left\{ \hspace*{-1mm} {\frac{1}{3}{{\log }_2}\hspace*{-1mm}\left( \hspace*{-0.5mm}{1 \hspace*{-1mm}+\hspace*{-1mm} {\mathrm{SINR}}_{{\mathrm{UE 1,UE 2}}}^{{{\mathrm{t}}_{\mathrm{1}}}}} \hspace*{-0.5mm}\right) \hspace*{-1mm}<\hspace*{-1mm} {R}_{{\mathrm{UE 2}}}^{{\mathrm{DL}}}} \hspace*{-1mm}\right\} \hspace*{-1mm}+\hspace*{-1mm}
\Pr \hspace*{-0.5mm}\left\{\hspace*{-1mm} {\frac{1}{3}{{\log }_2}\hspace*{-1mm}\left( \hspace*{-0.5mm}{1 \hspace*{-1mm}+\hspace*{-1mm} {\mathrm{SINR}}_{{\mathrm{UE 1,UE 2}}}^{{{\mathrm{t}}_{\mathrm{1}}}}} \hspace*{-0.5mm}\right) \hspace*{-1mm}\ge\hspace*{-1mm} {R}_{{\mathrm{UE 2}}}^{{\mathrm{DL}}},\frac{1}{3}{{\log }_2}\hspace*{-1mm}\left(\hspace*{-0.5mm} {1 \hspace*{-1mm}+\hspace*{-1mm} {\mathrm{SINR}}_{{\mathrm{UE 1}}}^{{{\mathrm{t}}_{\mathrm{1}}}}} \hspace*{-0.5mm}\right) \hspace*{-1mm}<\hspace*{-1mm} {R}_{{\mathrm{UE 1}}}^{{\mathrm{DL}}}} \hspace*{-1mm}\right\}, \label{OutageProbability_UE1Down}\\%[-0.5mm]
%%%%
\hspace*{-3mm}{\mathrm{P}}_{{\mathrm{out}}}^{{\mathrm{UE 2}},\;{\mathrm{DL}}} \hspace*{-1mm}=&\hspace*{-0.5mm} \Pr \left\{ {\frac{1}{3}{{\log }_2}\left( {1 \hspace*{-1mm}+\hspace*{-1mm} {\mathrm{SINR}}_{{\mathrm{UE 1}},{\mathrm{UE 2}}}^{{{\mathrm{t}}_{\mathrm{1}}}}} \right) < {R}_{{\mathrm{UE 2}}}^{{\mathrm{DL}}},\frac{1}{3}{{\log }_2}\left( {1 \hspace*{-1mm}+\hspace*{-1mm} {\mathrm{SINR}}_{{\mathrm{UE 2}}}^{{{\mathrm{t}}_{\mathrm{1}}}}} \right) < {R}_{{\mathrm{UE 2}}}^{{\mathrm{DL}}}} \right\} + \notag\\[-0.5mm]
&\Pr \left\{ {\frac{1}{3}{{\log }_2}\left( {1 \hspace*{-1mm}+\hspace*{-1mm} {\mathrm{SINR}}_{{\mathrm{UE 1}},{\mathrm{UE 2}}}^{{{\mathrm{t}}_{\mathrm{1}}}}} \right) \ge {R}_{{\mathrm{UE 2}}}^{{\mathrm{DL}}},\frac{1}{3}{{\log }_2}\left( {1 \hspace*{-1mm}+ \hspace*{-1mm} {\mathrm{SINR}}_{{\mathrm{UE 2}}}^{{{\mathrm{t}}_{\mathrm{1}}}{\mathrm{,}}{{\mathrm{t}}_{\mathrm{2}}}}} \right) < {R}_{{\mathrm{UE 2}}}^{{\mathrm{DL}}}} \right\}, \label{PoutUE2down}\\%[-0.5mm]
%%%%%%
\hspace*{-3mm}{\mathrm{P}}_{{\mathrm{out,\;t_2}}}^{{\mathrm{UE 1,\;UL}}} \hspace*{-1mm}=&\hspace*{-0.5mm} \Pr \left\{ {\frac{1}{3}{{\log }_2}\left( {1 \hspace*{-1mm}+\hspace*{-1mm} {\mathrm{SINR}}_{{\mathrm{BS,UE 1}}}^{{{\mathrm{t_2}}}}} \right) < {R}_{{\mathrm{UE 1}}}^{{\mathrm{UL}}}} \right\}, \label{OutageProbability_UE1UP}\\%[-0.5mm]
%%%%%
\hspace*{-3mm}{\mathrm{P}}_{{\mathrm{out,\;t_3}}}^{{\mathrm{UE 1,\;UL}}} \hspace*{-1mm}=&\hspace*{-0.5mm} \Pr \left\{ {{{\left| {{h_{{\mathrm{BS}},{\mathrm{UE1}}}}} \right|}^2} \hspace*{-1mm}\ge\hspace*{-1mm} {{\left| {{h_{{\mathrm{BS}},{\mathrm{UE2}}}}} \right|}^2},\frac{1}{3}{{\log }_2}\left( {1 \hspace*{-1mm}+\hspace*{-1mm} {\mathrm{SINR}}_{{\mathrm{BS}},{\mathrm{UE1}}}^{{{\mathrm{t}}_{\mathrm{3}}}}} \right) < R_{{\mathrm{UE1}}}^{{\mathrm{UL}}}} \right\} \notag\\%[-0.5mm]
&+ \Pr \left\{ {{{\left| {{h_{{\mathrm{BS}},{\mathrm{UE1}}}}} \right|}^2} \hspace*{-1mm}<\hspace*{-1mm} {{\left| {{h_{{\mathrm{BS}},{\mathrm{UE2}}}}} \right|}^2},\frac{1}{3}{{\log }_2}\left( {1 \hspace*{-1mm}+\hspace*{-1mm} {\mathrm{\overline{SINR}}}_{{\mathrm{BS}},{\mathrm{UE2}}}^{{{\mathrm{t}}_{\mathrm{3}}}}} \right) < R_{{\mathrm{UE2}}}^{{\mathrm{UL}}}} \right\} \notag\\%[-0.5mm]
&+ \Pr \left\{ {{{\left| {{h_{{\mathrm{BS}},{\mathrm{UE1}}}}} \right|}^2} \hspace*{-1mm}<\hspace*{-1mm} {{\left| {{h_{{\mathrm{BS}},{\mathrm{UE2}}}}} \right|}^2},\frac{1}{3}{{\log }_2}\left( {1 \hspace*{-1mm}+\hspace*{-1mm} {\mathrm{\overline{SINR}}}_{{\mathrm{BS}},{\mathrm{UE2}}}^{{{\mathrm{t}}_{\mathrm{3}}}}} \right) \ge R_{{\mathrm{UE2}}}^{{\mathrm{UL}}},\frac{1}{3}{{\log }_2}\left( {1 \hspace*{-1mm}+\hspace*{-1mm} {\mathrm{\overline{SINR}}}_{{\mathrm{BS}},{\mathrm{UE1}}}^{{{\mathrm{t}}_{\mathrm{3}}}}} \right) < R_{{\mathrm{UE1}}}^{{\mathrm{UL}}}} \right\}, \label{OutageProbability_UE1UP2}\\%[-0.5mm]
%%%%%%%
\hspace*{-3mm}{\mathrm{P}}_{{\mathrm{out}}}^{{\mathrm{UE 2,\;UL}}} \hspace*{-1mm}=&\hspace*{-0.5mm} \Pr \left\{ {{{\left| {{h_{{\mathrm{BS}},{\mathrm{UE2}}}}} \right|}^2} \hspace*{-1mm}\ge\hspace*{-1mm} {{\left| {{h_{{\mathrm{BS}},{\mathrm{UE1}}}}} \right|}^2},\frac{1}{3}{{\log }_2}\left( {1 \hspace*{-1mm}+\hspace*{-1mm} {\mathrm{\overline{SINR}}}_{{\mathrm{BS}},{\mathrm{UE2}}}^{{{\mathrm{t}}_{\mathrm{3}}}}} \right) < R_{{\mathrm{UE2}}}^{{\mathrm{UL}}}} \right\} \notag\\%[-0.5mm]
&+ \Pr \left\{ {{{\left| {{h_{{\mathrm{BS}},{\mathrm{UE2}}}}} \right|}^2} \hspace*{-1mm}<\hspace*{-1mm} {{\left| {{h_{{\mathrm{BS}},{\mathrm{UE1}}}}} \right|}^2},\frac{1}{3}{{\log }_2}\left( {1 \hspace*{-1mm}+\hspace*{-1mm} {\mathrm{{SINR}}}_{{\mathrm{BS}},{\mathrm{UE1}}}^{{{\mathrm{t}}_{\mathrm{3}}}}} \right) < R_{{\mathrm{UE1}}}^{{\mathrm{UL}}}} \right\} \notag\\%[-0.5mm]
&+ \Pr \left\{ {{{\left| {{h_{{\mathrm{BS}},{\mathrm{UE2}}}}} \right|}^2} \hspace*{-1mm}<\hspace*{-1mm} {{\left| {{h_{{\mathrm{BS}},{\mathrm{UE1}}}}} \right|}^2},\frac{1}{3}{{\log }_2}\left( {1 \hspace*{-1mm}+\hspace*{-1mm} {\mathrm{{SINR}}}_{{\mathrm{BS}},{\mathrm{UE1}}}^{{{\mathrm{t}}_{\mathrm{3}}}}} \right) \ge R_{{\mathrm{UE1}}}^{{\mathrm{UL}}},\frac{1}{3}{{\log }_2}\left( {1 \hspace*{-1mm}+\hspace*{-1mm} {\mathrm{{SINR}}}_{{\mathrm{BS}},{\mathrm{UE2}}}^{{{\mathrm{t}}_{\mathrm{3}}}}} \right) < R_{{\mathrm{UE2}}}^{{\mathrm{UL}}}} \right\}.\label{OutageProbability_UE2UP}
\end{align}
%\par
%\vspace*{-2mm}
%\noindent
% Restore the current equation number.
\setcounter{equation}{24}
% IEEE uses as a separator
\hrulefill
% The spacer can be tweaked to stop underfull vboxes.
\vspace*{-4mm}
\end{figure*}

The outage probability of UE 1 for downlink NOMA transmission has been derived in \cite{KimCDRT2015} as follows:
%\vspace*{-1mm}
\begin{align}
\hspace*{-2mm}{\mathrm{P}}_{{\mathrm{out}}}^{{\mathrm{UE1}},\;{\mathrm{DL}}} \hspace*{-1mm}=\hspace*{-1mm} \left\{ \hspace*{-2mm} \begin{array}{ll}
1 - {\exp( - \frac{{{\phi _1}}}{{{\beta _{{\mathrm{BS,UE1}}}}\rho}})}, &\hspace*{-2mm}\text{if} \; {\alpha_{{\mathrm{UE2}}}^{{\mathrm{t_1}}}} \hspace*{-1mm}-\hspace*{-1mm} {\alpha_{{\mathrm{UE1}}}^{{\mathrm{t_1}}} }\gamma_{{\mathrm{UE 2}}}^{\mathrm{DL}} \hspace*{-1mm}>\hspace*{-1mm} 0\\
1, &\hspace*{-2mm}\text{otherwise}\end{array} \right.
\end{align}
%\par
%\vspace*{-2mm}
%\noindent
where ${\phi _1} = \max \left\{ {\frac{{\gamma _{{\mathrm{UE2}}}^{{\mathrm{DL}}}}}{{\left( {\alpha _{{\mathrm{UE2}}}^{{{\mathrm{t}}_{\mathrm{1}}}} - \alpha _{{\mathrm{UE1}}}^{{{\mathrm{t}}_{\mathrm{1}}}}\gamma _{{\mathrm{UE2}}}^{{\mathrm{DL}}}} \right)}},\frac{{\gamma _{{\mathrm{UE1}}}^{{\mathrm{DL}}}}}{{\alpha _{{\mathrm{UE1}}}^{{{\mathrm{t}}_{\mathrm{1}}}}}}} \right\}$, $\gamma_{{\mathrm{UE 1}}}^{\mathrm{DL}} = 2^{3{R}_{{\mathrm{UE 1}}}^{\mathrm{DL}}}-1$, and $\gamma_{{\mathrm{UE 2}}}^{\mathrm{DL}} = 2^{3{R}_{{\mathrm{UE 2}}}^{\mathrm{DL}}}-1$. It is notable that the power allocation factors should satisfy ${\alpha_{{\mathrm{UE2}}}^{{\mathrm{t_1}}}} - {\alpha_{{\mathrm{UE1}}}^{{\mathrm{t_1}}} }\gamma_{{\mathrm{UE 2}}}^{\mathrm{DL}} > 0$, otherwise ${\mathrm{P}}_{{\mathrm{out}}}^{{\mathrm{UE1}},\;{\mathrm{DL}}}$ will always be one.

Based on \eqref{PoutUE2down}, the outage probability of UE 2 for downlink NOMA transmission is derived as \eqref{PoutUE2down2} at the top of next page,
\begin{figure*}[!t]
% ensure that we have normalsize text
\normalsize
% Store the current equation number.
\setcounter{mytempeqncnt}{\value{equation}}
% Set the equation number to one less than the one
% desired for the first equation here.
% The value here will have to changed if equations
% are added or removed prior to the place these
% equations are referenced in the main text.
\setcounter{equation}{25}
\begin{align}
{\mathrm{P}}_{{\mathrm{out}}}^{{\mathrm{UE 2}},\;{\mathrm{DL}}} =& \underbrace{\Pr \left\{ \frac{1}{3}{{\log }_2}\left( {1 \hspace*{-1mm}+\hspace*{-1mm} {\mathrm{SINR}}_{{\mathrm{UE 1}},{\mathrm{UE 2}}}^{{{\mathrm{t}}_{\mathrm{1}}}}} \right) < {R}_{{\mathrm{UE 2}}}^{{\mathrm{DL}}}\right\}}_{Q_1}\underbrace{\Pr \left\{\frac{1}{3}{{\log }_2}\left( {1 \hspace*{-1mm}+\hspace*{-1mm} {\mathrm{SINR}}_{{\mathrm{UE 2}},{\mathrm{UE 2}}}^{{{\mathrm{t}}_{\mathrm{1}}}}} \right) < {R}_{{\mathrm{UE 2}}}^{{\mathrm{DL}}} \right\}}_{Q_2} + \notag\\[-1mm]
&\underbrace{\Pr \left\{ \frac{1}{3}{{\log }_2}\left( {1 \hspace*{-1mm}+\hspace*{-1mm} {\mathrm{SINR}}_{{\mathrm{UE 1}},{\mathrm{UE 2}}}^{{{\mathrm{t}}_{\mathrm{1}}}}} \right) \ge {R}_{{\mathrm{UE 2}}}^{{\mathrm{DL}}}\right\}}_{1-Q_1}\underbrace{\Pr \left\{\frac{1}{3}{{\log }_2}\left( {1 \hspace*{-1mm}+ \hspace*{-1mm} {\mathrm{SINR}}_{{\mathrm{UE 2}},{\mathrm{UE 2}}}^{{{\mathrm{t}}_{\mathrm{1}}}{\mathrm{,}}{{\mathrm{t}}_{\mathrm{2}}}}} \right) < {R}_{{\mathrm{UE 2}}}^{{\mathrm{DL}}} \right\}}_{Q_3}.\label{PoutUE2down2}%[-0.5mm]
\end{align}
%\par
%\vspace*{-5mm}
%\noindent
% Restore the current equation number.
\setcounter{equation}{26}
% IEEE uses as a separator
\hrulefill
% The spacer can be tweaked to stop underfull vboxes.
\end{figure*}
where $Q_1$ and $Q_2$ can be easily obtained as
%\vspace*{-1mm}
\begin{align}
Q_1 = 1 - {\exp( - \frac{{{\phi _2}}}{{{\beta _{{\mathrm{BS,UE1}}}}\rho}})} \; \text{and} \;
Q_2 = 1 - {\exp( - \frac{{{\phi _2}}}{{{\beta _{{\mathrm{BS,UE2}}}}\rho}})},
\end{align}
%\par
%\vspace*{-1mm}
%\noindent
respectively, and $\phi _2 = \frac{{\gamma _{{\mathrm{UE2}}}^{{\mathrm{DL}}}}}{{\left( {\alpha _{{\mathrm{UE2}}}^{{{\mathrm{t}}_{\mathrm{1}}}} - \alpha _{{\mathrm{UE1}}}^{{{\mathrm{t}}_{\mathrm{1}}}}\gamma _{{\mathrm{UE2}}}^{{\mathrm{DL}}}} \right)}}$. Again, the prerequisite ${\alpha_{{\mathrm{UE2}}}^{{\mathrm{t_1}}}} - {\alpha_{{\mathrm{UE1}}}^{{\mathrm{t_1}}} }\gamma_{{\mathrm{UE 2}}}^{\mathrm{DL}} > 0$ should be satisfied, otherwise ${\mathrm{P}}_{{\mathrm{out}}}^{{\mathrm{UE2}},\;{\mathrm{DL}}}$ will always be one.

For $Q_3$, we first derive the distributions of ${\mathrm{SINR}_{{\mathrm{UE2}}}^{{{\mathrm{t}}_{\mathrm{1}}}}}$ and ${\mathrm{SINR}_{{\mathrm{UE2}}}^{{{\mathrm{t}}_{\mathrm{2}}}}}$, respectively, and then obtain $Q_3$ via the following integration:
%\vspace*{-2mm}
\begin{align}
{Q_3}{\mathrm{ = }}\hspace*{-3mm} {\mathop {\int\int}\limits_{{{\mathrm{SINR}}_{{\mathrm{UE2}}}^{{{\mathrm{t}}_1}} + {\mathrm{SINR}}_{{\mathrm{UE2}}}^{{{\mathrm{t}}_{\mathrm{2}}}} < \gamma _{{\mathrm{UE2}}}^{{\mathrm{DL}}}}} }\hspace*{-5mm}{{f_{{\mathrm{SINR}}_{{\mathrm{UE2}}}^{{{\mathrm{t}}_{\mathrm{2}}}}}}\mathop {\hspace*{-1mm}\left( x \right)} {f_{{\mathrm{SINR}}_{{\mathrm{UE2}}}^{{{\mathrm{t}}_1}}}}\hspace*{-1mm}\left( y \right)dy\hspace*{0.5mm}dx}.
\end{align}

The CDF of ${\mathrm{SINR}_{{\mathrm{UE2}}}^{{{\mathrm{t}}_{\mathrm{1}}}}}$ is defined as
%\vspace*{-1mm}
\begin{align}
{F_{{\mathrm{SINR}}_{{\mathrm{UE2}}}^{{{\mathrm{t}}_{\mathrm{1}}}}}}\left( x \right) &= \Pr \left\{ {{\mathrm{SINR}}_{{\mathrm{UE2}}}^{{{\mathrm{t}}_{\mathrm{1}}}} < x} \right\},
\end{align}
thereby, if $0<x<{\frac{{\alpha _{{\mathrm{UE2}}}^{{{\mathrm{t}}_{\mathrm{1}}}}}}{{\alpha _{{\mathrm{UE1}}}^{{{\mathrm{t}}_{\mathrm{1}}}}}}}$, the CDF and PDF of ${\mathrm{SINR}_{{\mathrm{UE2}}}^{{{\mathrm{t}}_{\mathrm{1}}}}}$ are given by
%\vspace*{-2mm}
\begin{align}
{F_{{\mathrm{SINR}}_{{\mathrm{UE2}}}^{{{\mathrm{t}}_{\mathrm{1}}}}}}\hspace*{-1.5mm}\left( x \right)\hspace*{-1mm} &= \hspace*{-1mm}{F_{{{\left| {{{{h}}_{{\mathrm{BS,UE2}}}}} \right|}^2}}}\hspace*{-1.5mm}\left(\hspace*{-1mm} {\frac{x}{\left({\alpha _{{\mathrm{UE2}}}^{{{\mathrm{t}}_{\mathrm{1}}}} \hspace*{-1mm}-\hspace*{-1mm} \alpha _{{\mathrm{UE1}}}^{{{\mathrm{t}}_1}}x}\right)\hspace*{-1mm}\rho}} \hspace*{-1mm}\right) \;\text{and}\\[-0.5mm]
{{f_{\mathrm{SINR}_{{\mathrm{UE2}}}^{{{\mathrm{t}}_{\mathrm{1}}}}}}\hspace*{-1.5mm}\left( x \right)} \hspace*{-1mm} &= \hspace*{-1mm} {f_{{{\left| {{{{h}}_{{\mathrm{BS,UE2}}}}} \right|}^2}}}\hspace*{-1.5mm}\left(\hspace*{-1mm}{\frac{x}{\left({\alpha _{{\mathrm{UE2}}}^{{{\mathrm{t}}_{\mathrm{1}}}} \hspace*{-1mm}-\hspace*{-1mm} \alpha _{{\mathrm{UE1}}}^{{{\mathrm{t}}_1}}x}\right)\hspace*{-1mm}\rho}}\hspace*{-1mm}\right)
\hspace*{-1mm}{\frac{\alpha _{{\mathrm{UE2}}}^{{{\mathrm{t}}_{\mathrm{1}}}}}{\left({\alpha _{{\mathrm{UE2}}}^{{{\mathrm{t}}_{\mathrm{1}}}} \hspace*{-1mm}-\hspace*{-1mm} \alpha _{{\mathrm{UE1}}}^{{{\mathrm{t}}_1}}x}\right)^2 \hspace*{-1mm}\rho}},
\end{align}
%\par
%\vspace*{-2mm}
%\noindent
respectively. Similarly, for $0<x<{\frac{{\alpha _{{\mathrm{UE2}}}^{{{\mathrm{t}}_{\mathrm{2}}}}}}{{\alpha _{{\mathrm{BS}}}^{{{\mathrm{t}}_{\mathrm{2}}}}}}}$, the CDF and PDF of ${\mathrm{SINR}_{{\mathrm{UE2}}}^{{{\mathrm{t}}_{\mathrm{2}}}}}$ can be obtained as
%\vspace*{-2mm}
\begin{align}
{F_{{\mathrm{SINR}}_{{\mathrm{UE2}}}^{{{\mathrm{t}}_{\mathrm{2}}}}}}\hspace*{-1.5mm}\left( x \right) \hspace*{-1mm} &= \hspace*{-1mm} {F_{{{\left| {{{{h}}_{{\mathrm{UE1,UE2}}}}} \right|}^2}}}\hspace*{-1.5mm}\left( \hspace*{-1mm} {\frac{x}{\left({\alpha _{{\mathrm{UE2}}}^{{{\mathrm{t}}_{\mathrm{2}}}} \hspace*{-1mm}-\hspace*{-1mm} \alpha _{{\mathrm{BS}}}^{{{\mathrm{t}}_2}}x}\right)\hspace*{-1mm}\rho}} \hspace*{-1mm}\right)\; \text{and}\\[-0.5mm]
{{f_{\mathrm{SINR}_{{\mathrm{UE2}}}^{{{\mathrm{t}}_{\mathrm{2}}}}}}\hspace*{-1.5mm}\left( x \right)} \hspace*{-1mm} &= \hspace*{-1mm} {f_{{{\left| {{{{h}}_{{\mathrm{UE1,UE2}}}}} \right|}^2}}}\hspace*{-1.5mm}\left(\hspace*{-1mm}{\frac{x}{\left({\alpha _{{\mathrm{UE2}}}^{{{\mathrm{t}}_{\mathrm{2}}}} \hspace*{-1mm}-\hspace*{-1mm} \alpha _{{\mathrm{BS}}}^{{{\mathrm{t}}_2}}x}\right)\hspace*{-1mm}\rho}}\hspace*{-1mm}\right)\hspace*{-1mm}{\frac{\alpha _{{\mathrm{UE2}}}^{{{\mathrm{t}}_{\mathrm{2}}}}}{\left({\alpha _{{\mathrm{UE2}}}^{{{\mathrm{t}}_{\mathrm{2}}}} \hspace*{-1mm}- \hspace*{-1mm}\alpha _{{\mathrm{BS}}}^{{{\mathrm{t}}_2}}x}\right)^2 \hspace*{-1mm}\rho}},
\end{align}
respectively. Then, $Q_3$ can be obtained by solving:
%\vspace*{-2mm}
\begin{align}
{Q_3}{\mathrm{ = \hspace*{-1mm}}}\int_0^{\phi_3} \hspace*{-1mm}{\int_0^{\gamma _{{\mathrm{UE2}}}^{\mathrm{DL}} - x} \hspace*{-1mm} {{f_{\mathrm{SINR}_{{\mathrm{UE2}}}^{{{\mathrm{t}}_{\mathrm{2}}}}}}\hspace*{-1mm}\left( x \right)} } {f_{\mathrm{SINR}_{{\mathrm{UE2}}}^{{{\mathrm{t}}_1}}}}\hspace*{-1mm}\left( y \right)dy \hspace*{0.5mm} dx,
\end{align}
%\par
%\vspace*{-2mm}
%\noindent
where $\phi_3 = \min \left( {\gamma _{{\mathrm{UE2}}}^{\mathrm{DL}},\frac{{\alpha _{{\mathrm{UE2}}}^{{{\mathrm{t}}_{\mathrm{2}}}}}}{{\alpha _{{\mathrm{BS}}}^{{{\mathrm{t}}_{\mathrm{2}}}}}}} \right)$.
%Note that the upper limits of integration are given as above since we should guarantee ${\alpha_{{\mathrm{UE2}}}^{{\mathrm{t_1}}}} - {\alpha_{{\mathrm{UE1}}}^{{\mathrm{t_1}}} }\gamma_{{\mathrm{UE 2}}}^{\mathrm{DL}} > 0$ but not need for ${\alpha_{{\mathrm{UE2}}}^{{\mathrm{t_2}}}} - {\alpha_{{\mathrm{BS}}}^{{\mathrm{t_2}}} }\gamma_{{\mathrm{UE 2}}}^{\mathrm{DL}} > 0$.

It is difficult to directly solve the above integration.
To obtain more insights from ${\mathrm{P}}_{{\mathrm{out}}}^{{\mathrm{UE 2}},\;{\mathrm{DL}}}$ in \eqref{PoutUE2down2}, we apply the Gauss-Chebyshev integration\cite{hildebrand1987introduction} to obtain $Q_3$ via a closed-form approximation as follows\footnote{The tightness of the adopted approximation will be verified in the simulation section.}:
%\vspace*{-2mm}
\begin{align}
\hspace*{-2mm}{Q_3} \hspace*{-0.75mm}\thickapprox \hspace*{-0.75mm} {F_{{\mathrm{SINR}}_{{\mathrm{UE2}}}^{{{\mathrm{t}}_{\mathrm{2}}}}}}\hspace*{-1.5mm}\left( {{\phi _3}} \right)\hspace*{-1mm} - \hspace*{-1mm}\frac{{\alpha _{{\mathrm{UE2}}}^{{{\mathrm{t}}_{\mathrm{2}}}}{\phi _3}}}{{2{\beta _{{\mathrm{UE1,UE2}}}}\rho }}\hspace*{-1mm}\sum\limits_{i = 1}^n {\hspace*{-0.5mm}\frac{\pi }{n}\hspace*{-0.5mm}\left| {\sin {\frac{{2i \hspace*{-1mm}-\hspace*{-1mm} 1}}{{2n}}\pi }} \right|} g\hspace*{-0.5mm}\left( {{l_i}} \right),
\end{align}
%\par
%\vspace*{-2mm}
%\noindent
where $n$ is the number of Gauss-Chebyshev integral approximation terms, ${l_i} = \frac{{{\phi _3}}}{2}+\frac{{{\phi _3}}}{2} \cos {\frac{{2i - 1}}{{2n}}\pi }$, and $g\left( x \right)$ is given by
\begin{align}
g\hspace*{-0.5mm}\left( x \right) \hspace*{-0.5mm} = & \frac{1}{{{{\left( {\alpha
_{{\mathrm{UE2}}}^{{{\mathrm{t}}_{\mathrm{2}}}} - \alpha
_{{\mathrm{BS}}}^{{{\mathrm{t}}_2}}x} \right)}^2}}}\hspace*{-0.7mm}\exp
\hspace*{-1mm}\left( \hspace*{-1mm} - \frac{x}{{\left( {\alpha
_{{\mathrm{UE2}}}^{{{\mathrm{t}}_{\mathrm{2}}}} - \alpha
_{{\mathrm{BS}}}^{{{\mathrm{t}}_2}}x} \right){\beta _{{\mathrm{UE1,UE2}}}}\rho }}
- \notag\right.\\
&\left.\frac{{\left( {\gamma _{{\mathrm{UE2}}}^{{\mathrm{DL}}}
- x} \right)}}{{\left( {\alpha _{{\mathrm{UE2}}}^{{{\mathrm{t}}_1}} - \alpha
_{{\mathrm{UE1}}}^{{{\mathrm{t}}_1}}\gamma _{{\mathrm{UE2}}}^{{\mathrm{DL}}} + \alpha
_{{\mathrm{UE1}}}^{{{\mathrm{t}}_1}}x} \right){\beta _{{\mathrm{BS,UE2}}}}\rho }}
\right).\label{GaussChebyshev}%[-0.5mm]
%%%%%%
\end{align}

Substitute $Q_1$, $Q_2$, and $Q_3$ into \eqref{PoutUE2down2}, if
${\alpha_{{\mathrm{UE2}}}^{{\mathrm{t_1}}}} \hspace*{-0.5mm}-\hspace*{-0.5mm} {\alpha_{{\mathrm{UE1}}}^{{\mathrm{t_1}}} }\gamma_{{\mathrm{UE 2}}}^{\mathrm{DL}} \hspace*{-1mm}>\hspace*{-1mm} 0$, the outage probability for the downlink transmission of UE 2 can be derived as \eqref{PoutUE2down3} at the top of next page.
\begin{figure*}[!t]
% ensure that we have normalsize text
\normalsize
% Store the current equation number.
\setcounter{mytempeqncnt}{\value{equation}}
% Set the equation number to one less than the one
% desired for the first equation here.
% The value here will have to changed if equations
% are added or removed prior to the place these
% equations are referenced in the main text.
\setcounter{equation}{36}\vspace*{-5mm}
\begin{align}
{\mathrm{P}}_{{\mathrm{out}}}^{{\mathrm{UE2}},\;{\mathrm{DL}}} = & \left( {1 - \exp
\left( { - \frac{{{\phi _2}}}{{{\beta _{{\mathrm{BS}},{\mathrm{UE1}}}}\rho }}}
\right)} \right)\left( {1 - \exp \left( { - \frac{{{\phi _2}}}{{{\beta
_{{\mathrm{BS}},{\mathrm{UE2}}}}\rho }}} \right)} \right)
- \exp \left( { - \frac{{{\phi _2}}}{{{\beta _{{\mathrm{BS}},{\mathrm{UE1}}}}\rho }}}
\right) \cdot \notag\\[-0.5mm]
&\left\{ {1 - \exp \left( { - \frac{{{\phi _3}}}{{{\beta
_{{\mathrm{UE1}},{\mathrm{UE2}}}}\left( {\alpha
_{{\mathrm{UE2}}}^{{{\mathrm{t}}_{\mathrm{2}}}} - \alpha
_{{\mathrm{BS}}}^{{{\mathrm{t}}_2}}{\phi _3}} \right)\rho }}} \right)
- \frac{{\alpha _{{\mathrm{UE2}}}^{{{\mathrm{t}}_{\mathrm{2}}}}{\phi _3}}}{{2{\beta
_{{\mathrm{UE1,UE2}}}}\rho }}\sum\limits_{i = 1}^n {\frac{\pi }{n}\left| {\sin \left(
{\frac{{2i - 1}}{{2n}}\pi } \right)} \right|} g\left( {{l_i}} \right)}
\right\}.\label{PoutUE2down3}%[-0.5mm]
\end{align}
% Restore the current equation number.
\setcounter{equation}{37}
% IEEE uses as a separator
\hrulefill
% The spacer can be tweaked to stop underfull vboxes.
\vspace*{-5mm}
\end{figure*}

In $\mathrm{t_2}$, since the interference of the weak user can be perfectly cancelled at the BS, the outage probability of UE 1 for uplink NOMA transmission can be easily obtained by
%\vspace*{-2mm}
\begin{equation}
{\mathrm{P}}_{{\mathrm{out}},\;{{\mathrm{t}}_{\mathrm{2}}}}^{{\mathrm{UE1}},\;{\mathrm{UL}}} = 1 - {\exp( - \frac{{\gamma _{{\mathrm{UE1}}}^{{\mathrm{UL}}}}}{{{\beta _{{\mathrm{BS,UE1}}}}\alpha _{{\mathrm{BS}}}^{{{\mathrm{t}}_{\mathrm{2}}}}\rho }})},\vspace*{-1mm}
\end{equation}
where $\gamma_{{\mathrm{UE 1}}}^{\mathrm{UL}} = 2^{3{R}_{{\mathrm{UE 1}}}^{\mathrm{UL}}}-1$.

For the uplink NOMA transmission phase, the outage probability is complicated since the integral area in \eqref{OutageProbability_UE1UP2} and \eqref{OutageProbability_UE2UP} depends on the target data rates of uplink transmissions of both users.
As a compromise solution, we focus on high data rate applications, e.g. ${{R}_{{\mathrm{UE 1}}}^{\mathrm{UL}}} > \frac{1}{3} \;\text{bit/s/Hz}$ and ${{R}_{{\mathrm{UE 2}}}^{\mathrm{UL}}} > \frac{1}{3} \;\text{bit/s/Hz}$. The closed-form outage probability of UE 1 for uplink NOMA transmission is derived in \eqref{PoutUE1up} at the top of next page,
wherein $\gamma_{{\mathrm{UE 2}}}^{\mathrm{UL}} = 2^{3{R}_{{\mathrm{UE 2}}}^{\mathrm{UL}}}-1$.
Note that $(a)$ in \eqref{PoutUE1up} holds when ${{R}_{{\mathrm{UE 1}}}^{\mathrm{UL}}} > \frac{1}{3} \;\text{bit/s/Hz}$ and ${{R}_{{\mathrm{UE 2}}}^{\mathrm{UL}}} > \frac{1}{3} \;\text{bit/s/Hz}$.
Similarly, the outage probability of the uplink transmission of UE 2 can be given by \eqref{PoutUE2up} at the top of next page.

\begin{figure*}[!t]
% ensure that we have normalsize text
\normalsize
% Store the current equation number.
\setcounter{mytempeqncnt}{\value{equation}}
% Set the equation number to one less than the one
% desired for the first equation here.
% The value here will have to changed if equations
% are added or removed prior to the place these
% equations are referenced in the main text.
\setcounter{equation}{38}
\begin{align}
{\mathrm{P}}_{{\mathrm{out}},\;{{\mathrm{t}}_{\mathrm{3}}}}^{{\mathrm{UE1}},\;{\mathrm{UL}}}
=&
\Pr \left\{ {{{\left| {{h_{{\mathrm{BS}},{\mathrm{UE1}}}}} \right|}^2} \ge {{\left|
{{h_{{\mathrm{BS}},{\mathrm{UE2}}}}} \right|}^2},{{\left|
{{h_{{\mathrm{BS}},{\mathrm{UE1}}}}} \right|}^2} \hspace*{-0.5mm}-\hspace*{-0.5mm} {{\left|
{{h_{{\mathrm{BS}},{\mathrm{UE2}}}}} \right|}^2}\gamma
_{{\mathrm{UE1}}}^{{\mathrm{UL}}} < \gamma _{{\mathrm{UE1}}}^{{\mathrm{UL}}}/\rho }
\right\} \notag\\%[-0.5mm]
&\hspace*{-0.5mm}+\hspace*{-0.5mm} \Pr \left\{ {{{\left| {{h_{{\mathrm{BS}},{\mathrm{UE1}}}}} \right|}^2} < {{\left|
{{h_{{\mathrm{BS}},{\mathrm{UE2}}}}} \right|}^2},{{\left|
{{h_{{\mathrm{BS}},{\mathrm{UE2}}}}} \right|}^2} \hspace*{-0.5mm}-\hspace*{-0.5mm} {{\left|
{{h_{{\mathrm{BS}},{\mathrm{UE1}}}}} \right|}^2}\gamma
_{{\mathrm{UE2}}}^{{\mathrm{UL}}} < \gamma _{{\mathrm{UE2}}}^{{\mathrm{UL}}}/\rho }
\right\} \notag\\%[-0.5mm]
&\hspace*{-0.5mm}+\hspace*{-0.5mm} \Pr \left\{ {{{\left| {{h_{{\mathrm{BS}},{\mathrm{UE1}}}}} \right|}^2} < {{\left|
{{h_{{\mathrm{BS}},{\mathrm{UE2}}}}} \right|}^2},{{\left|
{{h_{{\mathrm{BS}},{\mathrm{UE2}}}}} \right|}^2} \hspace*{-0.5mm}-\hspace*{-0.5mm} {{\left|
{{h_{{\mathrm{BS}},{\mathrm{UE1}}}}} \right|}^2}\gamma
_{{\mathrm{UE2}}}^{{\mathrm{UL}}} \ge \gamma _{{\mathrm{UE2}}}^{{\mathrm{UL}}}/\rho
,{{\left| {{h_{{\mathrm{BS}},{\mathrm{UE1}}}}} \right|}^2} < \gamma
_{{\mathrm{UE1}}}^{{\mathrm{UL}}}/\rho } \right\}\notag\\%[-0.5mm]
\mathop  = \limits^{(a)} & 1 \hspace*{-1mm}-\hspace*{-1mm} \Pr \hspace*{-0.5mm}\left\{
{{{\left| {{h_{{\mathrm{BS}},{\mathrm{UE1}}}}} \right|}^2}
\hspace*{-1mm}-\hspace*{-1mm} {{\left| {{h_{{\mathrm{BS}},{\mathrm{UE2}}}}}
\right|}^2}\gamma _{{\mathrm{UE1}}}^{{\mathrm{UL}}} \hspace*{-1mm}\ge\hspace*{-1mm}
\gamma _{{\mathrm{UE1}}}^{{\mathrm{UL}}}/\rho } \right\}
\hspace*{-1mm}-\hspace*{-1mm} \Pr \hspace*{-0.5mm}\left\{ {{{\left|
{{h_{{\mathrm{BS}},{\mathrm{UE2}}}}} \right|}^2} \hspace*{-1mm}-\hspace*{-1mm}
{{\left| {{h_{{\mathrm{BS}},{\mathrm{UE1}}}}} \right|}^2}\gamma
_{{\mathrm{UE2}}}^{{\mathrm{UL}}} \hspace*{-1mm}\ge\hspace*{-1mm} \gamma
_{{\mathrm{UE2}}}^{{\mathrm{UL}}}/\rho ,{{\left| {{h_{{\mathrm{BS}},{\mathrm{UE1}}}}}
\right|}^2} \hspace*{-1mm}<\hspace*{-1mm} \gamma
_{{\mathrm{UE1}}}^{{\mathrm{UL}}}/\rho } \right\}\notag\\%[-0.5mm]
= & 1 \hspace*{-1mm}-\hspace*{-1mm} \int_{\gamma
_{{\mathrm{UE1}}}^{{\mathrm{UL}}}/\rho }^{ \hspace*{-0.5mm}+\hspace*{-0.5mm} \infty } \hspace*{-1mm}{\int_0^{\frac{{ x
- \gamma _{{\mathrm{UE1}}}^{{\mathrm{UL}}}/\rho }}{{ \gamma
_{{\mathrm{UE1}}}^{{\mathrm{UL}}}}}} \hspace*{-3mm}{{f_{{{\left|
{{{\mathrm{h}}_{{\mathrm{BS}},{\mathrm{UE2}}}}} \right|}^2}}}\hspace*{-1mm}\left( y
\right)\hspace*{-1mm}{f_{{{\left| {{{\mathrm{h}}_{{\mathrm{BS}},{\mathrm{UE1}}}}}
\right|}^2}}}\hspace*{-1mm}\left( x \right)} } dy \hspace*{0.5mm}dx
\hspace*{-0.5mm}-\hspace*{-1mm} \int_{{\frac{{\gamma _{{\mathrm{UE2}}}^{{\mathrm{UL}}}
\hspace*{-0.5mm}+\hspace*{-0.5mm} \gamma _{{\mathrm{UE2}}}^{{\mathrm{UL}}}\gamma
_{{\mathrm{UE1}}}^{{\mathrm{UL}}}}}{{\rho }}}}^{ \hspace*{-0.5mm}+\hspace*{-0.5mm} \infty }
\hspace*{-1mm}{\int_{\gamma _{{\mathrm{UE1}}}^{{\mathrm{UL}}}/\rho }^{\frac{{x -
\gamma _{{\mathrm{UE2}}}^{{\mathrm{UL}}}/\rho }}{{\gamma
_{{\mathrm{UE2}}}^{{\mathrm{UL}}}}}} \hspace*{-3mm}{{f_{{{\left|
{{{\mathrm{h}}_{{\mathrm{BS}},{\mathrm{UE1}}}}} \right|}^2}}}\hspace*{-1mm}\left( y
\right)\hspace*{-1mm}{f_{{{\left| {{{\mathrm{h}}_{{\mathrm{BS}},{\mathrm{UE2}}}}}
\right|}^2}}}\hspace*{-1mm}\left( x \right)} } dy\hspace*{0.5mm}dx \notag\\%[-0.5mm]
= & 1 \hspace*{-0.5mm}- \hspace*{-0.5mm}\frac{{{\beta _{{\mathrm{BS,UE1}}}}{\,\exp(\hspace*{-0.5mm}-\hspace*{-0.5mm}\frac{{\gamma _{{\mathrm{UE1}}}^{{\mathrm{UL}}}}}{{{\beta
_{{\mathrm{BS,UE1}}}}\rho }})}}}{{\gamma
_{{\mathrm{UE1}}}^{{\mathrm{UL}}}{\beta _{{\mathrm{BS,UE2}}}}}\hspace*{-0.5mm}+\hspace*{-0.5mm}{\beta
_{{\mathrm{BS,UE1}}}}}\hspace*{-0.5mm}-\hspace*{-0.5mm}\frac{{{\beta _{{\mathrm{BS,UE2}}}}}}{{\gamma
_{{\mathrm{UE2}}}^{{\mathrm{UL}}}{\beta _{{\mathrm{BS,UE1}}}}}\hspace*{-0.5mm}+\hspace*{-0.5mm}{\beta
_{{\mathrm{BS,UE2}}}}}{\exp\Big(\hspace*{-0.5mm}-\hspace*{-0.5mm}\left( {\frac{{\gamma
_{{\mathrm{UE1}}}^{{\mathrm{UL}}}}}{{{\beta _{{\mathrm{BS,UE1}}}}\rho }} \hspace*{-0.5mm}+\hspace*{-0.5mm}
\frac{{\gamma _{{\mathrm{UE2}}}^{{\mathrm{UL}}} \hspace*{-0.5mm}+\hspace*{-0.5mm} \gamma
_{{\mathrm{UE2}}}^{{\mathrm{UL}}}\gamma _{{\mathrm{UE1}}}^{{\mathrm{UL}}}}}{{{\beta
_{{\mathrm{BS,UE2}}}}\rho }}} \right)\Big)},\label{PoutUE1up}\\%[-0.5mm]
{\mathrm{P}}_{{\mathrm{out}},\;{{\mathrm{t}}_{\mathrm{3}}}}^{{\mathrm{UE2}},\;{\mathrm{UL}}}
=& 1 \hspace*{-0.5mm}-\hspace*{-0.5mm} \frac{{{\beta _{{\mathrm{BS,UE2}}}}}{\,\exp\Big(\hspace*{-0.5mm}-\hspace*{-0.5mm}\frac{{\gamma _{{\mathrm{UE2}}}^{{\mathrm{UL}}}}}{{{\beta
_{{\mathrm{BS,UE2}}}}\rho }}\Big)}}{{\gamma
_{{\mathrm{UE2}}}^{{\mathrm{UL}}}{\beta _{{\mathrm{BS,UE1}}}}}\hspace*{-0.5mm}+\hspace*{-0.5mm}{\beta
_{{\mathrm{BS,UE2}}}}}\hspace*{-0.5mm}-\hspace*{-0.5mm}\frac{{{\beta _{{\mathrm{BS,UE1}}}}}}{{\gamma
_{{\mathrm{UE1}}}^{{\mathrm{UL}}}{\beta _{{\mathrm{BS,UE2}}}}}\hspace*{-0.5mm}+\hspace*{-0.5mm}{\beta
_{{\mathrm{BS,UE1}}}}}{\exp\Big(\hspace*{-0.5mm}-\hspace*{-0.5mm}\left( {\frac{{\gamma
_{{\mathrm{UE2}}}^{{\mathrm{UL}}}}}{{{\beta _{{\mathrm{BS,UE2}}}}\rho }} \hspace*{-0.5mm}+\hspace*{-0.5mm}
\frac{{\gamma _{{\mathrm{UE1}}}^{{\mathrm{UL}}} \hspace*{-0.5mm}+\hspace*{-0.5mm} \gamma
_{{\mathrm{UE1}}}^{{\mathrm{UL}}}\gamma _{{\mathrm{UE2}}}^{{\mathrm{UL}}}}}{{{\beta
_{{\mathrm{BS,UE1}}}}\rho }}} \right)\Big)}.\label{PoutUE2up}
\end{align}
% Restore the current equation number.
\setcounter{equation}{40}
% IEEE uses as a separator
\hrulefill
% The spacer can be tweaked to stop underfull vboxes.
\end{figure*}

Now, we analyze the diversity order for each link for our proposed scheme to obtain more insights into the system outage performance. The diversity order is defined as $d = \mathop {\lim }\limits_{\rho  \to \infty }  - \frac{{\log {\mathrm{P}}_{{\mathrm{out}}}}}{{\log \rho }}$ \cite{TseDMTradeoff} and the results are summarized in the following lemma.

\begin{Lem} By using the high SNR approximation, i.e., $1-\exp(-\frac{x}{\rho})\approx \frac{x}{\rho}$ \cite{Ding2015}, we obtain the diversity order for each communication link as:
\begin{align}
d_{{\mathrm{out}}}^{{\mathrm{UE1}},\;{\mathrm{DL}}} & = 1,\;
d_{{\mathrm{out}}}^{{\mathrm{UE2}},\;{\mathrm{DL}}} = 2,\;d_{{\mathrm{out,t_2}}}^{{\mathrm{UE1}},\;{\mathrm{UL}}} = 1,\label{DiversityOrder123}\\
d_{{\mathrm{out,t_3}}}^{{\mathrm{UE1}},\;{\mathrm{UL}}} &=0,\;\text{and}\;
d_{{\mathrm{out}}}^{{\mathrm{UE2}},\;{\mathrm{UL}}} = 0.\label{DiversityOrder5}
\end{align}
\end{Lem}

\begin{figure*}[!t]
% ensure that we have normalsize text
\normalsize
% Store the current equation number.
\setcounter{mytempeqncnt}{\value{equation}}
% Set the equation number to one less than the one
% desired for the first equation here.
% The value here will have to changed if equations
% are added or removed prior to the place these
% equations are referenced in the main text.
\setcounter{equation}{42}
\begin{align}
R=& \left(\hspace*{-0.5mm} {1 -
{\mathrm{P}}_{{\mathrm{out}}}^{{\mathrm{UE1}},\;{\mathrm{DL}}}} \hspace*{-0.5mm}
\right)R_{{\mathrm{UE1}}}^{{\mathrm{DL}}} +
\left(\hspace*{-0.5mm} {1 -
{\mathrm{P}}_{{\mathrm{out}}}^{{\mathrm{UE2}},\;{\mathrm{DL}}}} \hspace*{-0.5mm}
\right)R_{{\mathrm{UE2}}}^{{\mathrm{DL}}} +
\left(\hspace*{-0.5mm} {1 -
{\mathrm{P}}_{{\mathrm{out,}}\;{{\mathrm{t}}_2}}^{{\mathrm{UE1}},\;{\mathrm{UL}}}}
\hspace*{-0.5mm} \right)R_{{\mathrm{UE1}}}^{{\mathrm{UL}}}
+ \left( \hspace*{-0.5mm}{1 -
{\mathrm{P}}_{{\mathrm{out,}}\;{{\mathrm{t}}_3}}^{{\mathrm{UE1}},\;{\mathrm{UL}}}}
\hspace*{-0.5mm} \right)R_{{\mathrm{UE1}}}^{{\mathrm{UL}}}
+ \left( \hspace*{-0.5mm}{1 -
{\mathrm{P}}_{{\mathrm{out}}}^{{\mathrm{UE2}},\;{\mathrm{UL}}}} \hspace*{-0.5mm}
\right)R_{{\mathrm{UE2}}}^{{\mathrm{UL}}}.\label{OutageThroughput2}
\end{align}
\par
\vspace*{-2mm}
\noindent
% Restore the current equation number.
\setcounter{equation}{43}
% IEEE uses as a separator
\hrulefill
% The spacer can be tweaked to stop underfull vboxes.
\vspace*{-4mm}
\end{figure*}
The diversity order for the downlink transmission of UE 1 is one. Besides,
 the diversity order for the downlink transmission of UE 2 is two since there are two independent observations of the downlink messages of UE 2 in our proposed scheme. On the other hand, we obtain an uplink transmission for UE 1 with a diversity order of one via the superposition transmission during the cooperative phase. Interestingly, the diversity order for uplink NOMA transmission is zero, which implies that there is an error floor for the outage probability at high transmit SNR $\rho$.
This is due to  the lack of adaptive power control for uplink NOMA transmission leading to a  significant IUI  in the high transmit SNR regime\footnote{We note that the error floor inherently exists  in the uplink of cooperative NOMA schemes with fixed power allocation \cite{Tse2005,YangZhengNOMA}.
}.

On the other hand,  as all the nodes transmit their information at their fixed target data rates and the system throughput is determined by the outage probability. Therefore, to evaluate the spectral efficiency of our proposed scheme, we define the system outage throughput in \eqref{OutageThroughput2} at the top of this page.
\section{Simulation Results}
In this section, the performances of our proposed scheme are evaluated through simulations.
Without loss of generality, we assume that the variances of channel coefficient are ${\beta _{{\mathrm{BS,UE1}}}} = 1$, ${\beta _{{\mathrm{BS,UE2}}}} = 0.05$, and ${\beta _{{\mathrm{UE1,UE2}}}} = 0.8$.
The target data rates are $R_{{\mathrm{UE1}}}^{{\mathrm{DL}}} = R_{{\mathrm{UE2}}}^{{\mathrm{DL}}} = R_{{\mathrm{UE1}}}^{{\mathrm{UL}}} = R_{{\mathrm{UE2}}}^{{\mathrm{UL}}} = 1 \;\text{bit/s/Hz}$ and the power allocation factors are ${\alpha_{{\mathrm{UE1}}}^{{\mathrm{t_1}}} } = 0.05$, ${\alpha_{{\mathrm{UE2}}}^{{\mathrm{t_1}}} } = 0.95$,
${\alpha_{{\mathrm{BS}}}^{{\mathrm{t_2}}} } = 0.1$, and ${\alpha_{{\mathrm{UE2}}}^{{\mathrm{t_2}}} } = 0.9$.
The approximation parameter for Gauss-Chebyshev integration is set as $n = 100$.

\begin{figure}[t]
\centering\hspace*{-5mm}
\includegraphics[width=3.5 in]{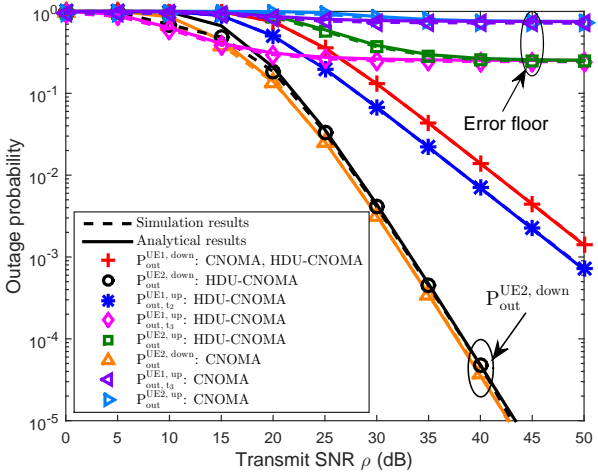}
\caption{Outage probability for the proposed HDU-CNOMA scheme and a conventional CNOMA scheme.}
\label{OutageProbability}
\end{figure}

\begin{figure}[t]
\centering\hspace*{-5mm}
\includegraphics[width=3.5 in]{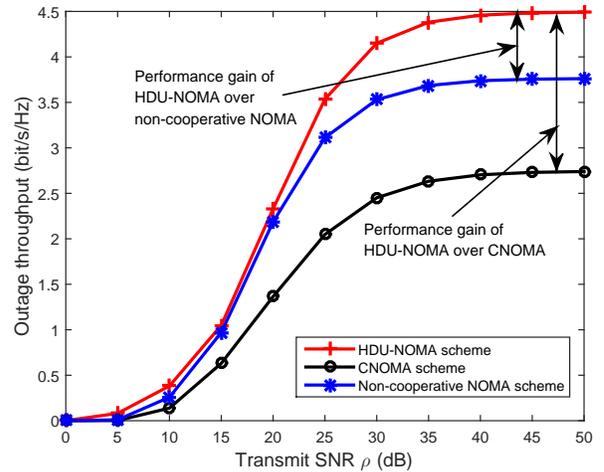}
\caption{Outage throughput (bits/s/Hz) for HDU-CNOMA scheme, conventional CNOMA scheme\cite{Ding2015}, and non-cooperative NOMA scheme.}
\label{OutageThroughput}
\end{figure}

Figure \ref{OutageProbability} illustrates the simulation results and analytical results for the outage probability of conventional CNOMA scheme and our proposed HDU-CNOMA scheme.
Note that the outage probability for ${\mathrm{P}}_{{\mathrm{out}}}^{{\mathrm{UE 1,\;DL}}}$ is the same for both CNOMA and HDU-CNOMA schemes.
It can be observed that our analytical results closely match with the simulation results, especially for the high SNR regime.
Compared to the CNOMA scheme, ${\mathrm{P}}_{{\mathrm{out}}}^{{\mathrm{UE 2,\;DL}}}$ of our proposed scheme is slightly higher due to the power loss in the cooperative phase. The gap on ${\mathrm{P}}_{{\mathrm{out}}}^{{\mathrm{UE 2,\;DL}}}$ between CNOMA and HDU-CNOMA can be further reduced by allocating a higher transmit power for UE 2 than that of the BS during the cooperative phase to maintain the signal reception reliability at UE 2.
On the other hand, although only a small faction of power is allocated for uplink transmission during $\mathrm{t_2}$ in HDU-CNOMA scheme, it has a lower outage probability than that of uplink NOMA transmissions in $\mathrm{t_3}$, especially for high SNR regime.
This is due to the fact that the side information $s_2$ assists the BS to cancel the interference in the superimposed signal transmitted during the cooperative phase.
For ${{R}_{{\mathrm{UE 1}}}^{\mathrm{UL}}} = {{R}_{{\mathrm{UE 2}}}^{\mathrm{UL}}} > \frac{1}{3} \;\text{bit/s/Hz}$, we can observe the error floor of ${\mathrm{P}}_{{\mathrm{out,t_3}}}^{{\mathrm{UE 1,\;UL}}}$ and ${\mathrm{P}}_{{\mathrm{out}}}^{{\mathrm{UE 2,\;UL}}}$ for both CNOMA and HDU-NOMA schemes, which validates our derivations in \eqref{DiversityOrder5}.
Also, it can be observed that our proposed scheme results in a lower error floor than that of the CNOMA scheme. This is because our proposed scheme exploit two time slots, $\mathrm{t_2}$ and $\mathrm{t_3}$, for UE 1 to transmit the target data rate ${{R}_{{\mathrm{UE 1}}}^{\mathrm{UL}}}$ while CNOMA only transmits in $\mathrm{t_3}$.

Figure \ref{OutageThroughput} depicts the outage throughput for all the schemes shown in Figure \ref{TransmissionIllustration}.
It can be observed that our proposed scheme achieve the largest outage throughput. In particular, the proposed scheme offers  substantial performance gains over the two baseline schemes in the moderate to high SNR regime.
Although the superimposed transmission of the proposed scheme during cooperative phase slightly degrades the received signal quality  at UE 2,  the performance gain brought by  the  extra interference-free uplink transmission of UE 1 outweighs the performance loss at UE 2 which  increases the overall system outage throughput. In contrast, the CNOMA scheme has a lowest outage throughput due to the following two reasons. First, compared to the proposed HDU-CNOMA scheme, the CNOMA scheme does not fully exploit the degrees of freedom in the system for uplink and downlink communications. Second, compared to the non-cooperative NOMA scheme,  the performance of  the CNOMA scheme relies on the existence of   short range communication between the strong
user and the weak user \cite{Ding2015} which does not always exist in practical systems.

\section{Conclusion}
In this paper, a novel HDU-CNOMA scheme was proposed to increase the spectral efficiency and to achieve a better tradeoff between signal reception reliability and spectral efficiency for cooperative NOMA systems.
Particularly, the cooperative transmission and uplink transmission were integrated during the cooperative phase, and the side information at the BS was utilized to obtain an additional interference-free uplink transmission.
To evaluate the performance of our proposed scheme, we analyzed the corresponding outage probability, diversity order, and system outage throughput.
Simulations were conducted to verify our analytical results.
With only a slightly performance degradation on the signal reception reliability at the weak user, our proposed scheme provides a substantial improvement on system spectral efficiency over a conventional cooperative NOMA scheme and a non-cooperative NOMA scheme.

\end{document}